# Looping and reconfiguration dynamics of a flexible chain with internal friction

Nairhita Samanta, Jayanta Ghosh and Rajarshi Chakrabarti*

*Department of Chemistry, Indian Institute of Technology Bombay,*

*Mumbai, Powai 400076, \*E-mail: rajarshi@chem.iitb.ac.in*

In recent past, experiments and simulations have suggested that apart from the solvent friction, friction arising from the protein itself plays an important role in protein folding by affecting the intra-chain loop formation dynamics. This friction is termed as internal friction in the literature. Using a flexible Gaussian chain with internal friction we analyze the intra-chain reconfiguration and loop formation times for all three topology classes namely end-to-end, end-to-interior and interior-to-interior. In a nutshell, bypassing expensive simulations we show how simple models like that of Rouse and Zimm can support the single molecule experiment and computer simulation results on intra-chain diffusion coefficients, looping time and even can predict the effects of tail length on the looping time.

## I. INTRODUCTION

Loop formation between two ends of a polymer chain has extensively been studied by experimentalists and polymer theorists. Theoretical models, computer simulations [1-29] hand in hand with experiments on long chain biomolecules such as proteins, DNA [30-35] have been quite successful in shedding light on the time scale of loop formation and its dependence on solvent viscosity [6], solvent quality [26, 27] and the chain length [3, 21]. Most of the experimental studies have dealt with end-to-end loops, i.e. between two end monomers of a chain and the theoretical models initially proposed were also for the end-to-end looping. But it is obvious that the end-to-end looping is only a sub-class of the looping dynamics as processes like protein aggregation [36], polypeptide and protein folding involves contact formation between any two segments of a protein molecule not necessarily between the two ends. Recent studies on end-to-interior looping dynamics also have shown that the tail ends have profound effect on the dynamics [20, 34, 37-39]. The effect of tail length on the looping or reconfiguration time of a chain has been attributed to the excluded volume effect of the chain. One can understand this by viewing the tail to intervene the looping process by mere obstructing. So, at first glance it appears to be solely due to the excluded volume effects.

Another very recently explored aspect is the effect of internal or solvent independent friction on looping dynamics [37, 40-44]. This internal or dry friction can be viewed as the intrinsic resistance of a polymer chain to undergo a conformational change and is independent of solvent viscosity, operative even at very low viscosity of the solvent. It is believed that hydrogen bonding, intra-chain collisions, dihedral angle rotation and clashes of chemical groups [39] contribute to internal friction. But a systematic understanding of internal friction is lacking. Other than some coarse-grained simulations [41, 45, 46] not much of theoretical investigations have taken place in this direction which could help in understanding the origin of internal friction. But experimentally it is now well established that internal friction plays an important role in protein folding kinetics. For example Cellmer et. al [40] used nanosecond laser T-jump to measure the viscosity dependence of the folding dynamics of the villain subdomain and found a major contribution to the dynamics from internal friction. Several studies using single molecule FRET by Schuler's [42, 43] group have shown the existence of non-vanishing reconfiguration time of the cold shock protein and spectrin domains even at extrapolated zero viscosity of the solvent. They found internal friction to have an additive contribution to the reconfiguration time of unfolded protein. Another important observation was the dependence of the magnitude of the internal friction on the compactness of the protein [43]. More compact is the protein higher the contribution from internal friction. This is understandable as different parts of an expanded proteins would be quite apart to interact through weak forces which is believed to contribute to internal friction. On the other hand, in a more compact situation different parts of a protein are close enough to interact through weak forces such as hydrogen bonds, also more likely to undergo intra-chain collisions and hence contributing more to the internal friction and would result in higher internal friction. Based on their experimental observations they proposed a relation between the reconfiguration time $\tau_r$ and the time scale associated with the internal friction $\tau_{int}$ as follows $\tau_r = \tau_{int} + \frac{\eta}{\eta_0}\tau_s(\eta_0)$, where $\eta_0$ is the viscosity of water, $\tau_s$ is the solvent dependent relaxation time, $\eta$ is the viscosity of the solvent. Obviously this relation assumed internal friction is independent of solvent viscosity. At zeroth level this assumption is not too bad though internal friction would depend on the solvent if not on the solvent viscosity. Other than the solvent viscosity, solvent quality and the concentration of the solvent would definitely affect the extent of solvation of different parts of the protein, so as the friction between these different parts of the protein molecule. Thus to calculate the reconfiguration time of a chain at different denaturant concentrations, one should use different values of $\tau_{int}$. Such entanglement of extent of solvation and internal friction has also been observed by Schulz et. al [41] in case of disordered glycine-serine and $\alpha$-helix forming alanine peptides. The above relation confirms that at the extrapolated zero solvent viscosity, the reconfiguration time is non-zero and equal to the time scale for internal friction. The above relation is also in consistent with a model such as a Rouse or Zimm chain with internal friction where the internal friction appears as an additive constant to each normal modes of the chain. Thus, $\tilde{\tau}_{p,R} = {\tau_R}/{p^2} + \tau_{int}$ and for the $\tilde{\tau}_{p,Z} = {\tau_Z}/{p^{3/2}} + \tau_{int}$ for the Rouse and Zimm chain respectively [47, 48], with $\tau_R$ and $\tau_Z$ are the slowest relaxation times for the corresponding chains. More recently using nuclear magnetic resonance and laser photolysis Yasin et. al [44] have shown that the degree of non-bonded atom interactions, which is a measure of protein compactness largely influence the extent of internal friction.

In this paper, we address two issues. Firstly we extensively investigate the looping dynamics of a Rouse and pre-averaged Zimm chain while considering all three topology classes, namely (i) interior to interior, (ii) end to interior and (iii) end to end (Fig. (1)). Without considering excluded volume effects we show how the chain length, tail length can have similar influence on the reconfiguration time for a Gaussian chain. We also find similar behaviour of the looping time when Wilemski-Fixman [1] theory is applied with the assumption that the chain is also piecewise Gaussian. Secondly we introduce internal friction and analyze its effect on the dynamics. We make a comparison between the contribution of excluded volume interactions [37] and internal friction on looping dynamics and find that the internal friction is playing similar role as that of excluded volume interactions and can slow down the dynamics as effectively as excluded volume interactions. Thus it is practically impossible to unentangle the effects of internal friction and excluded volume interactions.

## II. POLYMER MODEL

The simplest possible dynamical model for a single polymer is Rouse model [49, 50], which does not consider hydrodynamic and excluded volume interactions. The dynamics of a single Rouse chain is described by the following equation

$$\zeta \frac{\partial R_n(t)}{\partial t} = k \frac{\partial^2 R_n(t)}{\partial n^2} + f(n,t) \tag{1}$$

Where $R_n(t)$ is the position vector at time $t$, $n$ represent the monomer position and can take any value between 0 to $N$, $N+1$ is the total number of monomers present in the chain. $\zeta$ is the solvent viscosity and the spring constant $k = \frac{3k_B T}{b^2}$ where $b$ is the kuhn length, $f(n,t)$ is the random force acting on the polymer emerging from the random collisions between the polymer and solvent molecules with moments

$$\langle f(n,t) \rangle = 0, \qquad \langle f_\alpha(n,t_1) f_\beta(m,t_2) \rangle = 2\zeta k_B T \delta_{\alpha\beta} \delta(n-m) \delta(t_1 - t_2)$$

The above equation can be solved using normal mode analysis where the normal modes follow the equation

$$\zeta_{p,R} \frac{dX_p(t)}{dt} = -k_{p,R} X_p(t) + f_p(t) \tag{2}$$

Here, $\zeta_{p,R} = 2N\zeta$ for $p > 0$ and $\zeta_0 = N\zeta$. The relaxation time of each mode is $\tau_{p,R} = \frac{\zeta_{p,R}}{k_p} = \frac{\tau_R}{p^2}$, $k_p = \frac{6\pi^2 k_B T p^2}{Nb^2}$ and $\tau_R = \frac{N^2 \zeta b^2}{3\pi^2 k_B T}$ is the slowest relaxation time or the Rouse time. It can be shown that in $\theta$ solvent under pre-averaged hydrodynamic interaction the modes of a polymer behave like Rouse modes [49] with $\zeta_{p,Z} = \zeta \sqrt{\frac{\pi N p}{3}}$. The relaxation time become $\tau_{p,Z} = \frac{\zeta_{p,Z}}{k_p} = \frac{\tau_Z}{p^{3/2}}$. Similarly $\tau_Z = \frac{N^{3/2} \zeta b^2}{6\sqrt{3} \pi^{3/2} k_B T}$ the slowest relaxation time for the Zimm chain in $\theta$ solvent, called Zimm time.

In recent studies a model which can be decomposed into normal modes has been used to study the effect of internal or dry friction on polymer dynamics [20, 42, 43, 47, 48, 51] where the time scale associated with the internal friction adds to the relaxation time for the each mode. Thus formally one can write, $\tilde{\tau}_{p,Z} = \tau_{p,Z} + \tau_{int}$ for the Rouse chain with internal friction and $\tilde{\tau}_{p,Z} = \tau_{p,Z} + \tau_{int}$ for the pre-averaged Zimm chain in $\theta$ solvent with internal friction, where $\tau_{p,R} = \tau_R/p^2$ and $\tau_{p,Z} = \tau_Z/p^{3/2}$ and $\tau_{int} = \zeta_{int}/k$, $\zeta_{int}$ being the internal friction coefficient.

Also notice in absence as well as in presence of internal friction, the normal mode time correlation function has the following structure [47, 48, 52]

$$\langle X_{p\alpha}(0) X_{q\beta}(t) \rangle = \frac{k_B T}{k_p} \delta_{pq} \delta_{\alpha\beta} \exp(-t/\tau_p) \quad (3)$$

Here $\tau_p$ is $\tau_{p,R}$ for a Rouse chain, $\tilde{\tau}_{p,R}$ for a Rouse chain with internal friction, $\tau_{p,Z}$ for a Zimm chain and $\tilde{\tau}_{p,Z}$ for a Zimm chain with internal friction.

### III. RECONFIGURATION TIME AND MEAN SQUARE DISPLACEMENT

The reconfiguration time $\tau_{mn}$ for a polymer chain is defined as [37, 48]

$$\tau_{mn} = \int_0^\infty dt \left( \frac{\phi_{mn}(t)}{\phi_{mn}(0)} \right) \quad (4)$$

Where, $\phi_{mn}(t)$ is the time correlation function for the position vectors $\boldsymbol{R}_{mn}(t) = \boldsymbol{R}_m(t) - \boldsymbol{R}_n(t)$ and formally written as

$$\phi_{mn}(t) = \langle \boldsymbol{R}_{mn}(0) \cdot \boldsymbol{R}_{mn}(t) \rangle \quad (5)$$

Naturally faster decay of the correlation function $\phi_{mn}$ would result shorter reconfiguration time. To calculate the time correlation function $< \boldsymbol{R}_{mn}(0) \cdot \boldsymbol{R}_{mn}(t) >$ we make use of the normal mode analysis [49, 52] and in the $N \to \infty$ limit get

$$\boldsymbol{R}_{mn}(t) = 2 \sum_{p=1}^\infty \boldsymbol{X}_p(t) \left( \cos\left(\frac{p\pi m}{N}\right) - \cos\left(\frac{p\pi n}{N}\right) \right) \quad (6)$$

Next we make use of Eq. (3) to arrive at

$$\phi_{mn}(t) = \langle \boldsymbol{R}_{mn}(0) \cdot \boldsymbol{R}_{mn}(t) \rangle = 12 k_B T \sum_{p=1}^\infty k_p^{-1} \exp(-t/\tau_p) \left( \cos\left(\frac{p\pi m}{N}\right) - \cos\left(\frac{p\pi n}{N}\right) \right)^2 \quad (7)$$

The reconfiguration time as defined in Eq. (4) would then read as

$$\tau_{mn} = \frac{\sum_{p=1}^{\infty} \tau_p \left(\cos\left(\frac{p\pi m}{N}\right) - \cos\left(\frac{p\pi n}{N}\right)\right)^2}{\sum_{p=1}^{\infty} \frac{1}{p^2}\left(\cos\left(\frac{p\pi m}{N}\right) - \cos\left(\frac{p\pi n}{N}\right)\right)^2} \tag{8}$$

Next we find the analytically exact expression for the mean square displacement (MSD). MSD for the vector connecting $m$ and $n$ monomers is $\langle(\boldsymbol{R}_{mn}(t) - \boldsymbol{R}_{mn}(0))^2\rangle$ which can formally be written as

$$\langle(\boldsymbol{R}_{mn}(t) - \boldsymbol{R}_{mn}(0))^2\rangle = 4\sum_{p=1}^{\infty} \frac{Nb^2}{p^2\pi^2}\left(\cos\left(\frac{p\pi m}{N}\right) - \cos\left(\frac{p\pi n}{N}\right)\right)^2 \left(1 - \exp(-t/\tau_p)\right) \tag{9}$$

Putting $m = 0$ and $n = N$ in the above expression one gets the MSD for the end to end vector [21]. Also the above expression for the MSD can be expressed in terms of $\phi_{mn}(t)$.

$$\langle(\boldsymbol{R}_{mn}(t) - \boldsymbol{R}_{mn}(0))^2\rangle = 2(\phi_{mn}(0) - \phi_{mn}(t)) \tag{10}$$

On rearranging $\phi_{mn}(t)$ can be expressed as

$$\phi_{mn}(t) = \left(\phi_{mn}(0) - \frac{\langle(\boldsymbol{R}_{mn}(t) - \boldsymbol{R}_{mn}(0))^2\rangle}{2}\right) \tag{11}$$

The above expressions, for $\phi_{mn}(t)$, $\tau_{mn}$ and MSD are formally exact and hold for the Rouse and the Zimm chain with or without internal friction as long as the chain is Gaussian. For example, in case of a Zimm chain with internal friction, one should substitute $\tau_p$ by $\tilde{\tau}_{p,Z}$.

### IV. LOOP FORMATION TIME

One of the theoretical models to calculate the loop formation time between two ends of a long chain molecule is due to Wilemski and Fixman (WF) [1]. WF theory gives following expression for the loop formation time $\tau_{loop}$ between two ends of a Gaussian chain.

$$\tau_{loop} = \int_0^\infty dt \left(\frac{C(t)}{C(\infty)} - 1\right) \qquad (12)$$

where, C(t) is the sink-sink correlation function, defined as

$$C(t) = \int d\mathbf{R} \int d\mathbf{R_0}\, S(\mathbf{R}) G(\mathbf{R}, t|\mathbf{R_0}, t_0 = 0) S(\mathbf{R_0}) P(\mathbf{R_0}) \qquad (13)$$

In the above expression, $G(\mathbf{R}, t|\mathbf{R_0}, t_0 = 0)$ is the conditional probability that a chain with end-to-end distance at $R_0$ at time $t = t_0 = 0$ has an end-to-end distance $R$ at time $t$. It is assumed that the chain was initially in equilibrium at $t = t_0 = 0$ and has the equilibrium probability distribution, $P(R_0) = (\frac{3}{2\pi N b^2})^{3/2} \exp[-\frac{3R_0^2}{2Nb^2}]$. S(R) is the sink function [53-55] which takes care of the loop formation between two ends of the chain and is a function of only end to end distance. The above expression is a limiting case where the strength of the sink function is taken to be infinity.

For a Gaussian chain the conditional probability or the Green function is well known

$$G(\mathbf{R}, t|\mathbf{R_0}, t_0 = 0) = \left(\frac{3}{2\pi \langle R^2 \rangle_{eq}}\right)^{\frac{3}{2}} \left(\frac{1}{(1-\phi^2(t))^{\frac{3}{2}}}\right) \exp\left[-\frac{3(\mathbf{R}-\phi(t)\mathbf{R_0})^2}{2\langle R^2 \rangle_{eq}(1-\phi^2(t))}\right] \qquad (14)$$

where

$$\phi(t) = \frac{\langle \mathbf{R}(t).\mathbf{R}(0) \rangle_{eq}}{\langle R^2 \rangle_{eq}} \qquad (15)$$

is the normalized end-to-end vector correlation function of the chain. For a Gaussian chain the above formula can be rewritten in terms of the normal modes of the chain.

$$\phi(t) = \sum_{p=odd} \frac{8}{p^2 \pi^2} \exp[-t/\tau_p] \qquad (16)$$

The above equation is valid for a Rouse, Zimm chain in $\theta$ solvent and for a Rouse or Zimm chain with internal friction as long as the chain is Gaussian [33]. Also notice, $\phi(t)$ is nothing but $\phi_{mn}(t)$ with $m = 0$ and $n = N$.

Now here we generalize this theory for the loop-formation between any two monomers of the same chain. As long as the chain is Gaussian putting a sink between $m$ and $n$ monomers would result the following expression for the loop formation time

$$\tau_{mn,loop} = \int_0^\infty dt \left(\frac{C_{mn}(t)}{C_{mn}(\infty)} - 1\right) \quad (17)$$

where, $C_{mn}(t)$ is the sink-sink correlation function, defined as

$$C_{mn}(t) = \int d\mathbf{R}_{mn} \int d\mathbf{R}_{mn,0}\, S(\mathbf{R}_{mn})G(\mathbf{R}_{mn}, t|\mathbf{R}_{mn,0}, t_0 = 0)S(\mathbf{R}_{mn,0})P(\mathbf{R}_{mn,0}) \quad (18)$$

Similarly in the above expression, $G(\mathbf{R}_{mn}, t|\mathbf{R}_{mn,0}, t_0 = 0)$ is the conditional probability that a chain with $|m - n|$ at $R_{mn,0}$ at time $t = t_0 = 0$ has a distance $R_{mn}$ at time $t$. It is assumed that the chain was initially in equilibrium at $t = t_0 = 0$ and has the equilibrium probability distribution, $P(R_{mn,0}) = \left(\frac{3}{2\pi\langle R_{mn}^2\rangle_{eq}}\right)^{3/2}\exp\left[-\frac{3R_{mn,0}^2}{2\langle R_{mn}^2\rangle_{eq}}\right]$. $S(R_{mn})$ is the sink function which takes care of the loop formation between two monomers of the chain and is a function of the separation between the two monomers, $m$ and $n$, i.e. $|m - n|$. As before the above expression is a limiting case where the strength of the sink function is taken to be infinity.

Since we have assumed the chain to be piecewise Gaussian, the conditional probability or the Green's function is also Gaussian

$$G(\mathbf{R}_{mn}, t|\mathbf{R}_{mn,0}, t_0 = 0) = \left(\frac{3}{2\pi\langle R_{mn}^2\rangle_{eq}}\right)^{\frac{3}{2}} \left(\frac{1}{(1-\phi_{mn}^2(t))^{\frac{3}{2}}}\right) \exp\left[-\frac{3(\mathbf{R}_{mn} - \phi_{mn}(t)\mathbf{R}_{mn,0})^2}{2\langle R_{mn}^2\rangle_{eq}(1-\phi_{mn}^2(t))}\right]$$

$$(19)$$

where $\langle R_{mn}^2\rangle_{eq} = |m - n|b^2$

It is possible with a proper choice of the sink function to analytically derive an exact expression for the sink-sink correlation function (Eq. (13)). For example, if one chooses the sink function $S(R_{mn})$ to be a radial delta function $\delta(R_{mn} - a)$ [53, 54], where $a$ is the capture radius then the closed form expression for the loop formation time is given by [5, 25]

$$\tau_{mn,loop} = \int_0^\infty dt \left( \frac{\exp[-2\chi_0 \phi_{mn}^2(t)/(1-\phi_{mn}^2(t))]\sinh[(2\chi_0 \phi_{mn}(t))/(1-\phi_{mn}^2(t))]}{(2\chi_0 \phi_{mn}(t))\sqrt{1-\phi_{mn}^2(t)}} - 1 \right) \quad (20)$$

Where

$$\chi_0 = \frac{3a^2}{2\langle R_{mn}^2 \rangle_{eq}} \quad (21)$$

We would also like to mention here that assuming the polymer to have a Gaussian distribution is actually goes well with the experimental observations. For example, in the work by Möglich et. al [33], use of Gaussian model for a chain with internal friction has been well justified. We are quoting from them: "TTET experiments on end-to-end loop formation in the same peptide has previously shown kinetics in the Gaussian chain limit both in water and in presence of high GdmCl concentrations, indicating that this approximation is valid."

## V. RESULTS AND DISCUSSIONS

To calculate the reconfiguration time defined in Eq. (4) we first compute the correlation function defined in Eq. (5) and put this back in Eq. (4). A plot of $\phi_{mn}(t)$ for the Rouse chain and Rouse chain with internal friction (inset) are shown in Fig. (2). Here we consider three distinct topology classes, (i) end-to-end, (ii) end-to-interior and (iii) interior-to-interior (Fig. (1)). At a first look one would expect to see faster decay of $\phi_{mn}(t)$ with increase in the separation between the monomers, $|m-n|$ and slowest in case of end-to-end topology. But surprisingly it is not the case as can be seen from Fig. (2), where we consider two topology classes (i) and (ii) and compare. We choose a polymer with 67 monomers (N + 1 = 67), with parameters, $b = 3.8 \times 10^{-10}, \zeta = 9.42 \times 10^{-12} kgs^{-1}$ in consistence with the viscosity of water at $T = 300K$. This choice of parameters is motivated by the experiments done by Schuler's group [35]. We choose, $n = 66$ and a series of values of $m = 1, 9, 20, 21, 33$ and $m = 0$. It is clearly not the end-to-end correlation but rather $\phi_{mn}(t)$ with an inner monomer $m = 9$ and the other end with $n = 66$ shows slowest decay. But on choosing two interior monomers, $m = 33$ and $n = 66$ faster decay is observed. Thus, there is one primary factor, the separation between the monomers along the chain, $|m-n|$ which makes the correlation function $\phi_{mn}(t)$ to decay faster as the separation is reduced. But there is a secondary factor which actually favors faster decay of $\phi_{mn}(t)$s decay very fast and practically indistinguishable and the time scale associated with the movements of the long tail is too large to intervene. Another observation is the slight decrease of the looping time with large tail length. This is in consistence with what Doucet et. al [20] have observed in their simulations while considering excluded volume effects. Obviously in our case this cannot be due to excluded volume interactions. However this decrease is rather weak and more visible with internal friction.

Before we conclude we would like to comment on the possibility of including the effect of weak interactions on internal friction. Ideally one should carry out computer simulations to quantify the effect of weak interaction on internal friction. But still within our model one can incorporate this effect by invoking an ansatz, $\zeta_{int} = \zeta_{int,0} f(n_b)$, where $\zeta_{int,0}$ is the internal friction in absence of any weak attractive interactions, $n_b$ is the number of monomers interacting through non-bonded weak forces and should depend on the quality of the solvent [56], in other words it is a measure of the extent of weak interactions. Poorer the solvent higher the value of $n_b$ as then the intra-chain attractions are favored where beyond nearest neighbor interactions contribute, $f(n_b)$ is the multiplicative correction factor to the internal friction due to weak interactions. A simple choice of the functional form of $f(n_b)$ would be $f(n_b) = (c_0 + c_1 n_b + c_2 n_b^2 + \dots)$. Obviously with a choice of coefficients $c_0 = 1$, and $c_i = 0$ for $i > 0$, one gets back internal friction where only nearest neighbour interactions contribute $\zeta_{int} = \zeta_{int,0}$. On the other hand with the choice $c_i = 1$ for all $i$s, taking up to the second order term and taking $n_b = 2$ one gets a 7 times increase in the internal friction, $\zeta_{int} = 7\zeta_{int,0}$. Thus in our calculation use of successively higher values of the internal friction (from Fig. (4) to Fig. (13)) can be interpreted as the effect of weak attraction being taken care of.

## VI. CONCLUSIONS

In this paper we calculate reconfiguration time and loop formation time between two monomers of a single chain within Rouse and Zimm descriptions and analyze the effect of internal friction. Even without invoking excluded volume effects we observe similar features as observed in computer simulations explicitly taking excluded volume effects into consideration. It seems internal friction is playing similar role as that of excluded volume interaction. This is not very surprising as internal friction arises because of the resistance of a monomer to move with respect to the next monomer [48] and this resistance is a measure of weak interactions between neighboring monomers connected through chemical bonds as well as of excluded volume effects. In other words internal friction has considerable contributions from exclude volume interactions. But quantifying the amount of contribution from excluded volume interactions to internal friction requires computer simulations. A natural choice would be to carry out a systematic simulation of real proteins to find out the correlation between the excluded volume parameter(s) and the internal friction coefficient. Also our formulation of internal friction is purely dry as it does not depend on solvent viscosity, but protein molecules have a large number of solvent molecules around and thus the friction arises within the molecule should depend on the extent of solvation [45]. Computer simulation would also be extremely useful to address this issue. Another direction would be to analyze the effect of non-equilibrium initial configuration [24] of chain on looping and reconfiguration dynamics. But a more challenging task ahead is to incorporate internal friction to the theories of protein folding or any rate process involving long chain molecules. Routinely Kramers like theories are being used to calculate the rate of such processes [38, 57, 58]. Very recently in the context of enzyme reactions a Kramers theory with internal friction has been proposed [59]. But it is completely ad hoc lacking any first principle derivation since the internal friction has no corresponding fluctuation dissipation theorem. This invokes another question whether a Kramers like description is possible in presence of internal friction.

## VII. ACKNOWLEDGEMENT

N. S., J. G. and R. C. thank IRCC IIT Bombay for funding (Project Code: 12IRCCSG046).

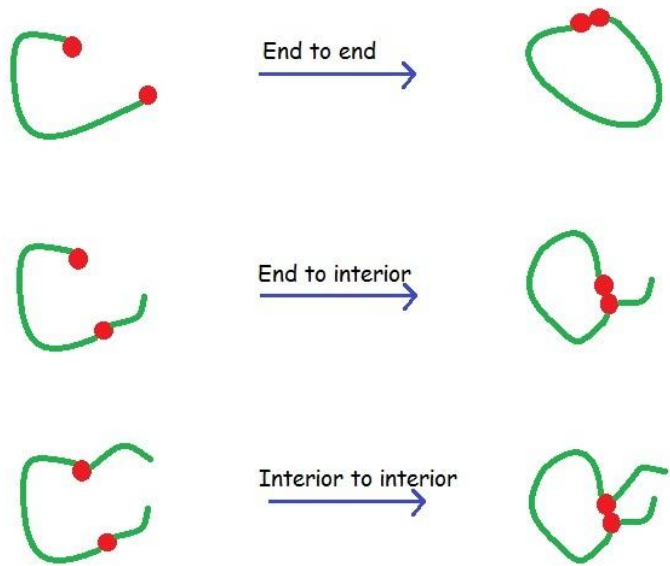

FIG. 1: Three topology classes in intra-chain loop formation.

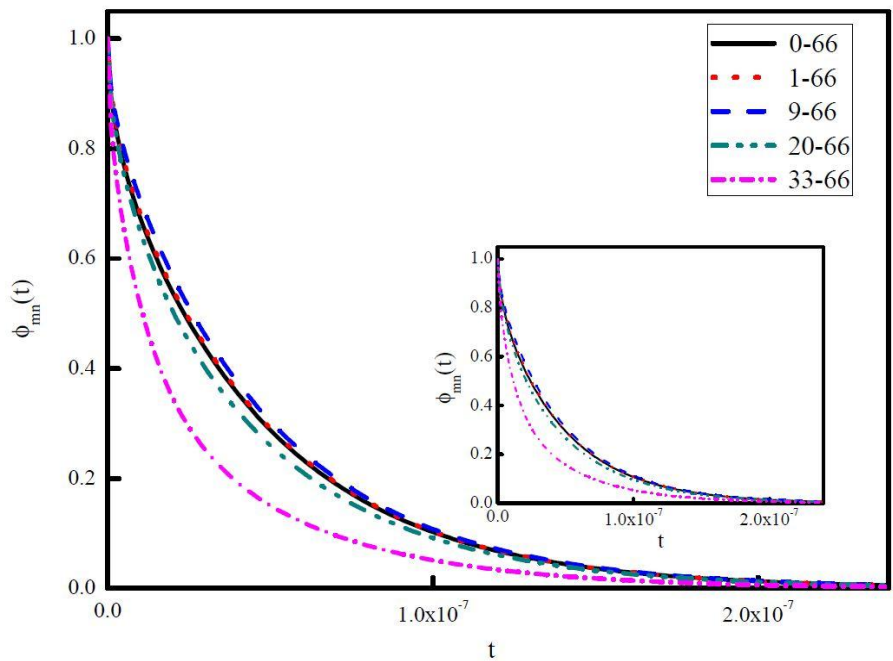

FIG. 2: $\phi_{mn}(t)$ vs $t$ for the Rouse chain and Rouse chain with internal friction $\zeta_{int} = 10\zeta$ (inset) for different set of $m$ and $n$ values (see legend).

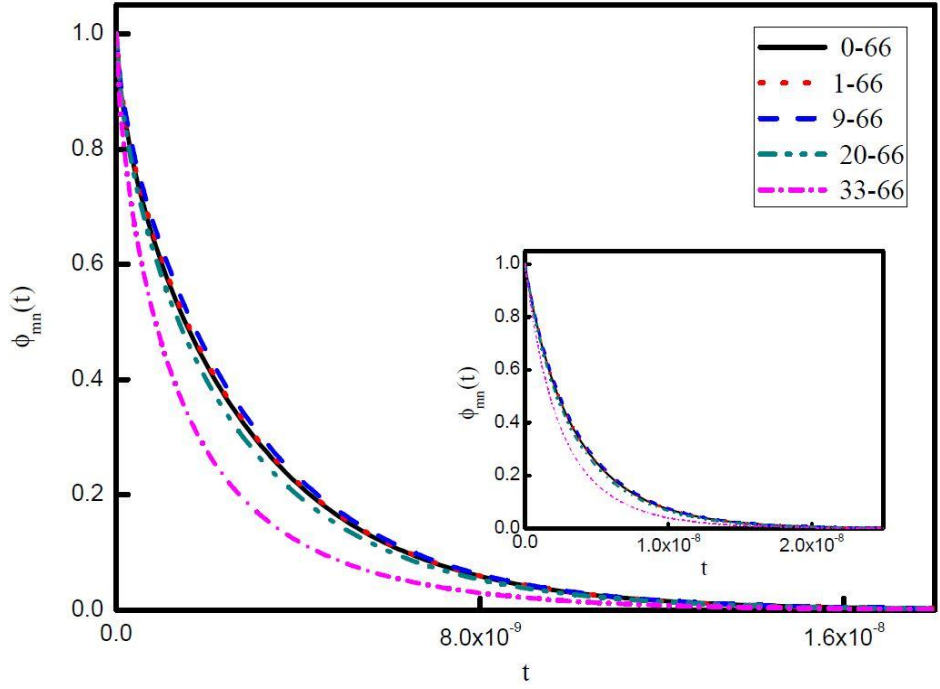

FIG. 3: $\phi_{mn}(t)$ vs $t$ for the Zimm chain and Zimm chain with internal friction $\zeta_{int} = 10\zeta$ (inset) for different set of $m$ and $n$ values (see legend).

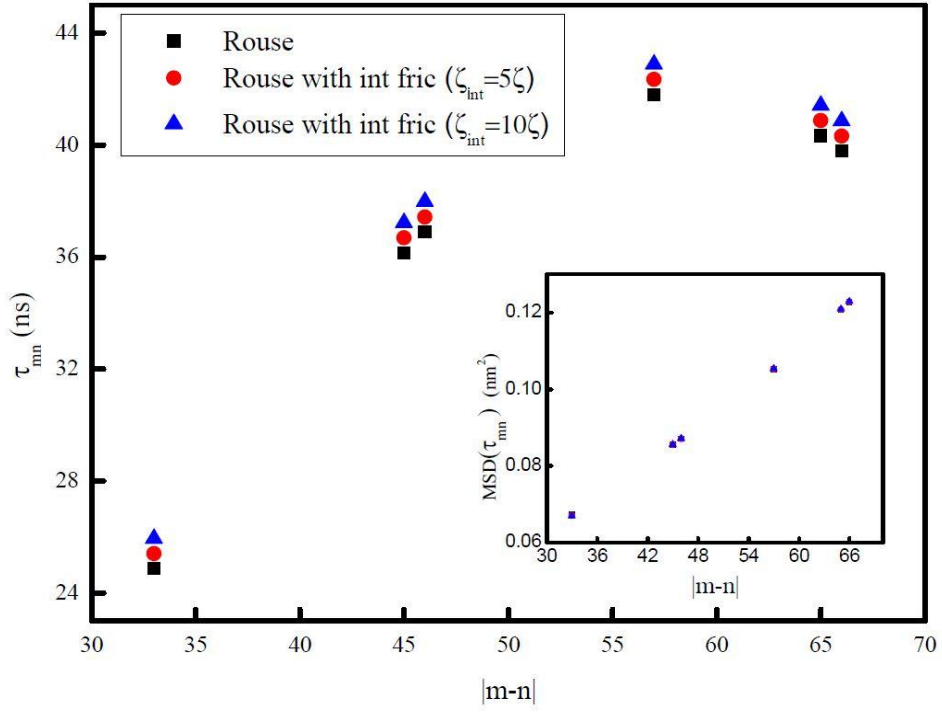

FIG. 4: $\tau_{mn}$ and MSD at $t = \tau_{mn}$ (inset) for the Rouse chain and Rouse chain with internal friction for the topology classes end-to-interior and end-to-end.

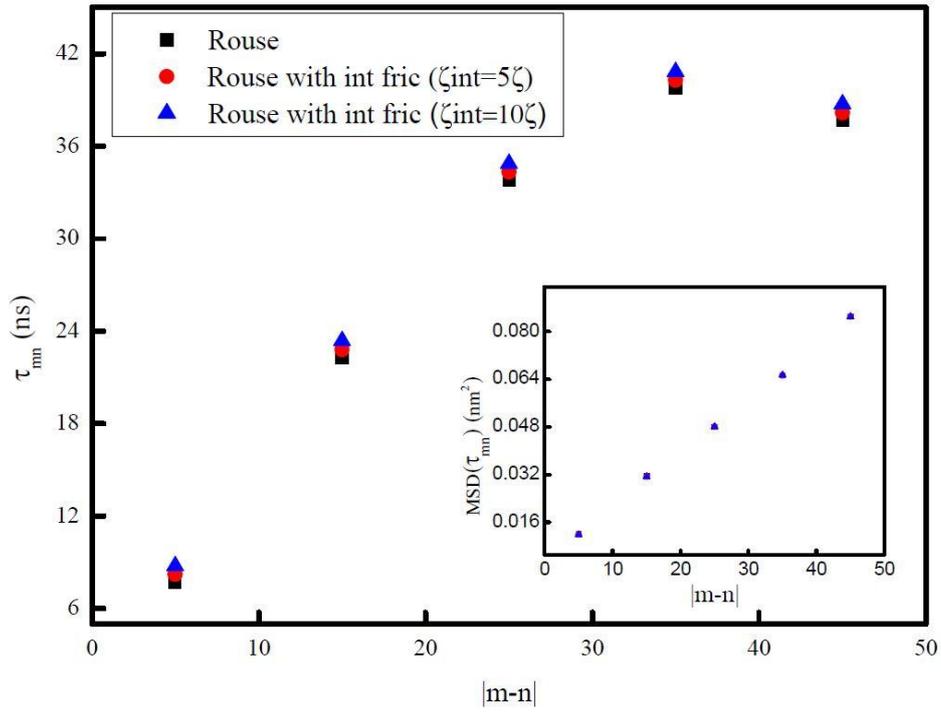

FIG. 5: $\tau_{mn}$ and MSD at $t = \tau_{mn}$ (inset) for the Rouse chain and Rouse chain with internal friction for the topology class interior-to-interior.

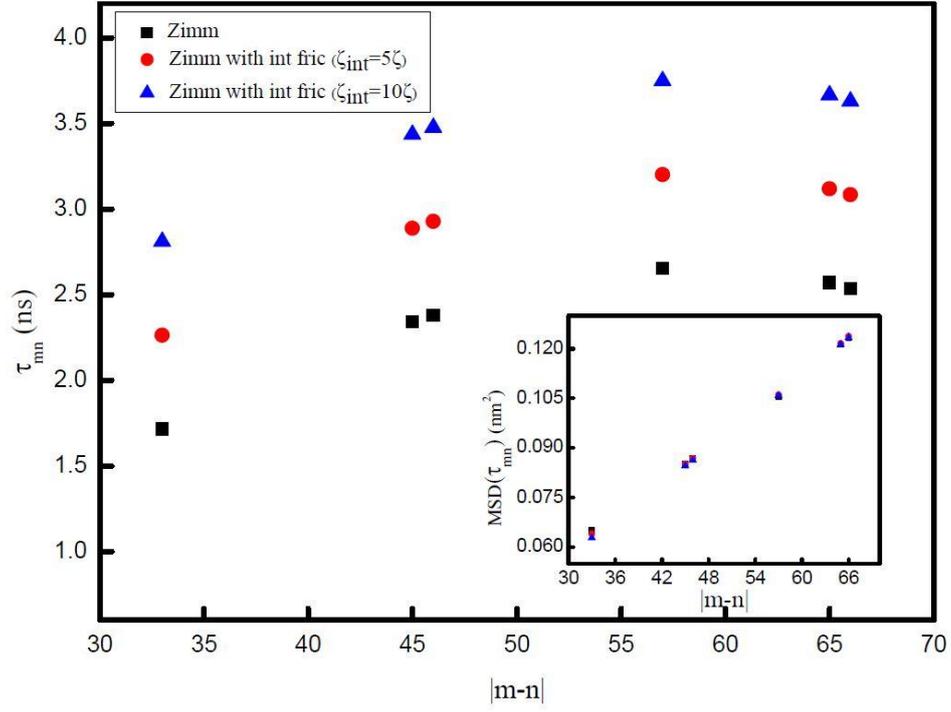

FIG. 6: $\tau_{mn}$ and MSD at $t = \tau_{mn}$ (inset) for the Zimm chain and Zimm chain with internal friction for the topology classes end-to-interior and end-to-end.

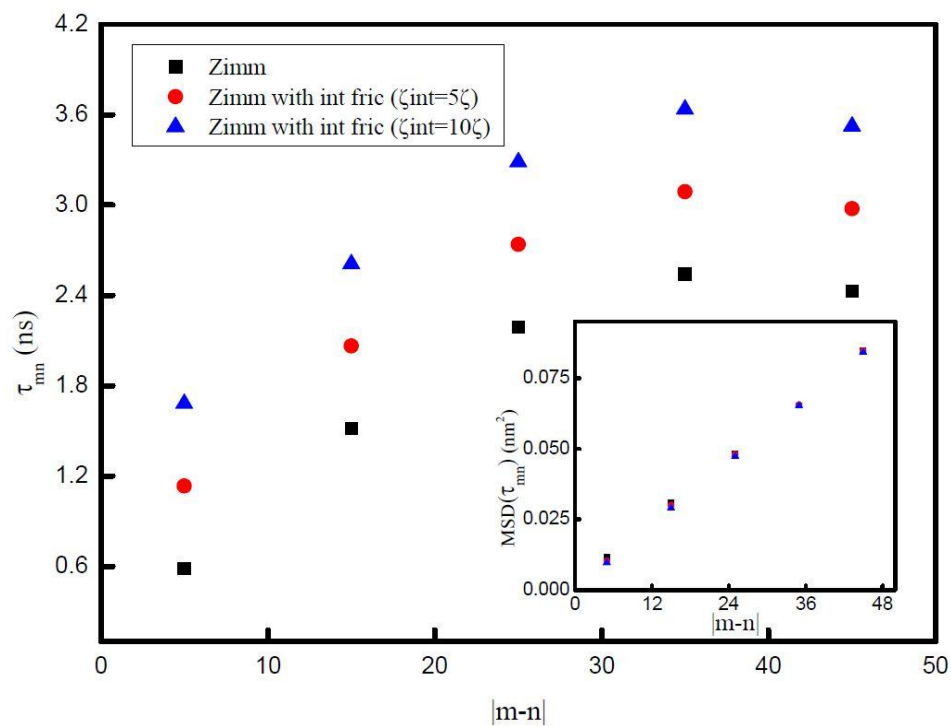

FIG. 7: $\tau_{mn}$ and MSD at $t = \tau_{mn}$ (inset) for the Zimm chain and Zimm chain with internal friction for the topology class interior-to-interior.

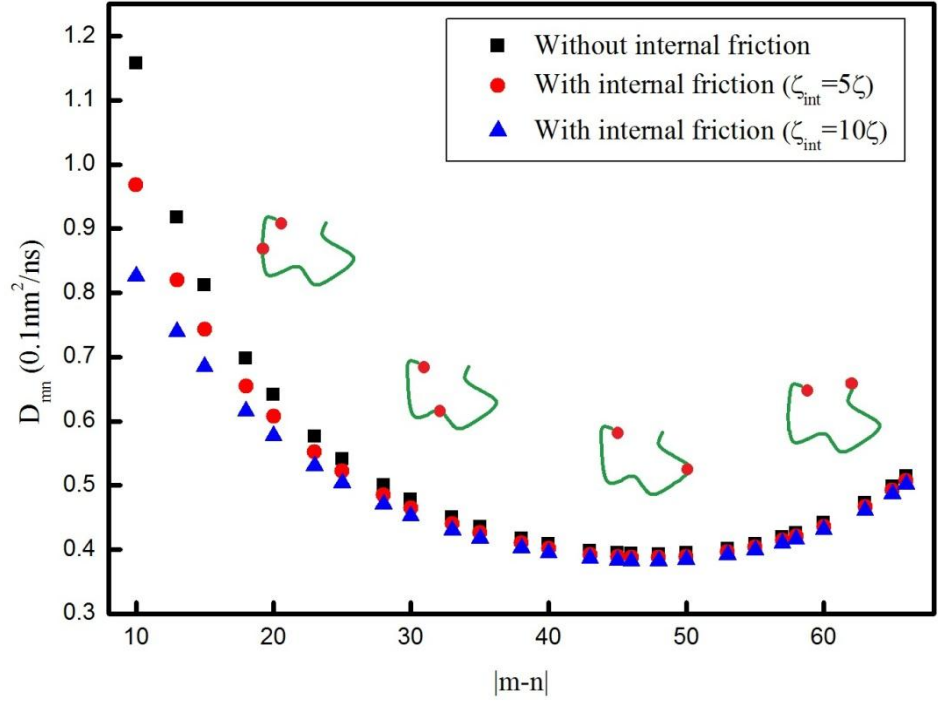

FIG. 8: Intra-chain diffusion coefficient, $D_{mn}$ vs intra-chain separation $|m-n|$ for the Rouse chain. Cartoons of polymers tentatively show different topology classes considered.

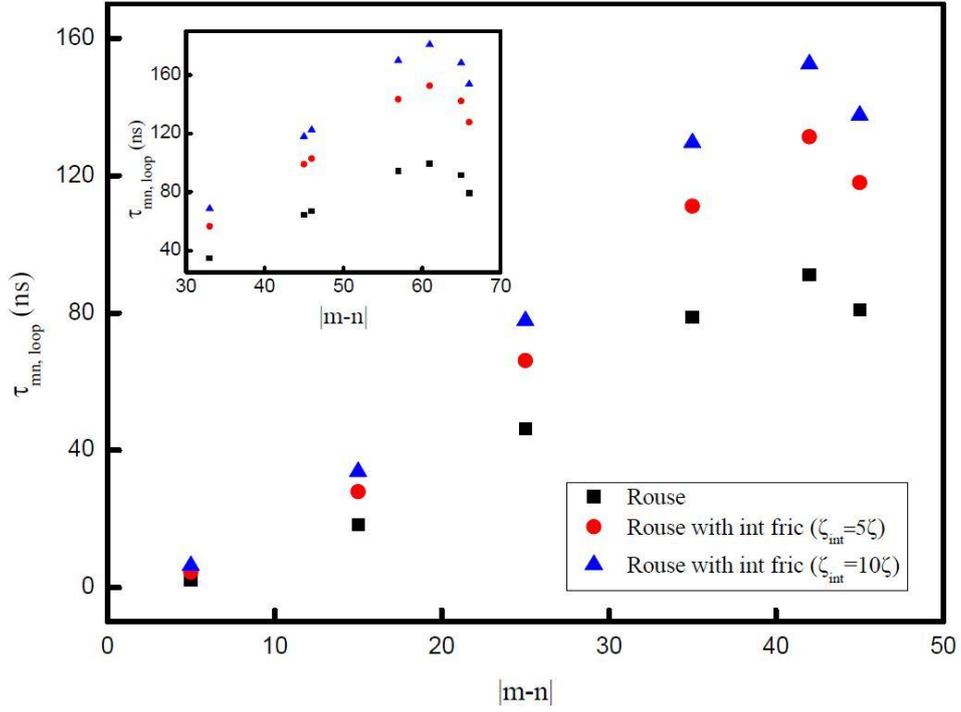

FIG. 9: $\tau_{mn,loop}$ for the Rouse chain and Rouse chain with internal friction for the topology classes interior-interior and end-to-interior (inset).

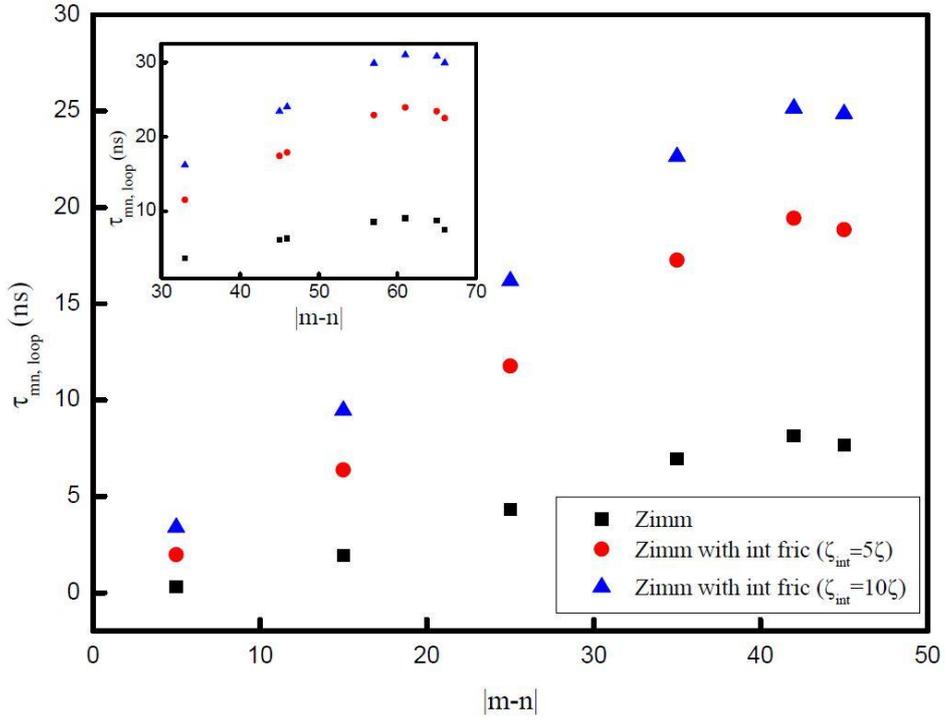

FIG. 10: $\tau_{mn,loop}$ for the Zimm chain and Zimm chain with internal friction for the topology classes interior-interior and end-to-interior (inset).

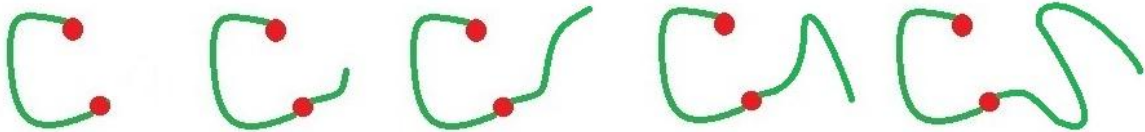

FIG. 11: To analyze the effect of tail length on looping, we start from the end-to-end configuration and then a tail is added to one end and successively the tail length is increased.

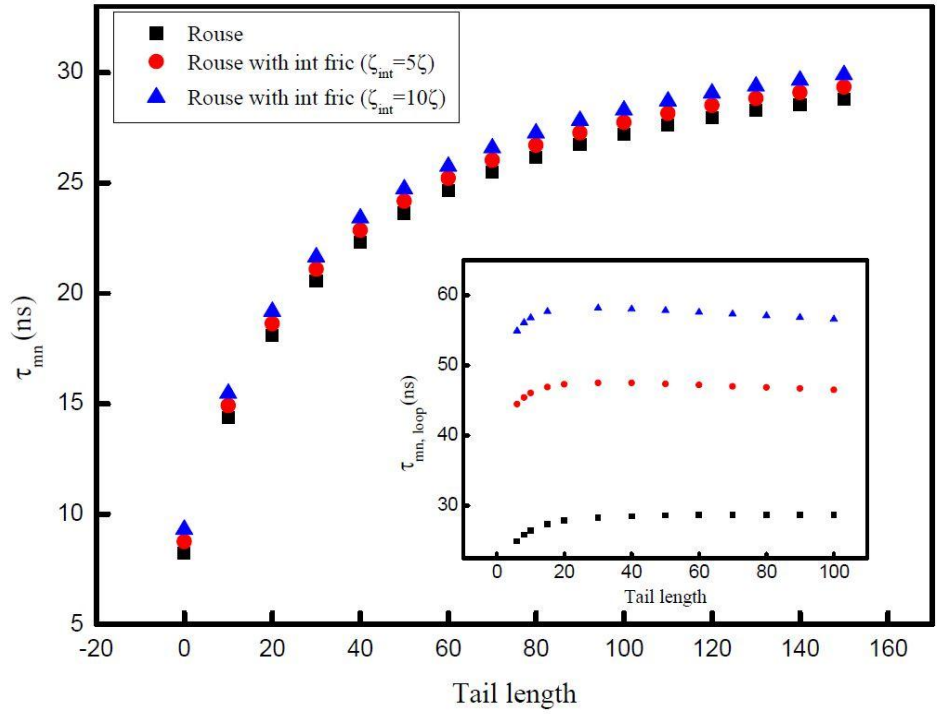

FIG. 12: $\tau_{mn}$ and $\tau_{mn,loop}$ (inset) vs tail length for the Rouse chain and Rouse chain with internal friction, for the topology classes end-to-end and end-to-interior.

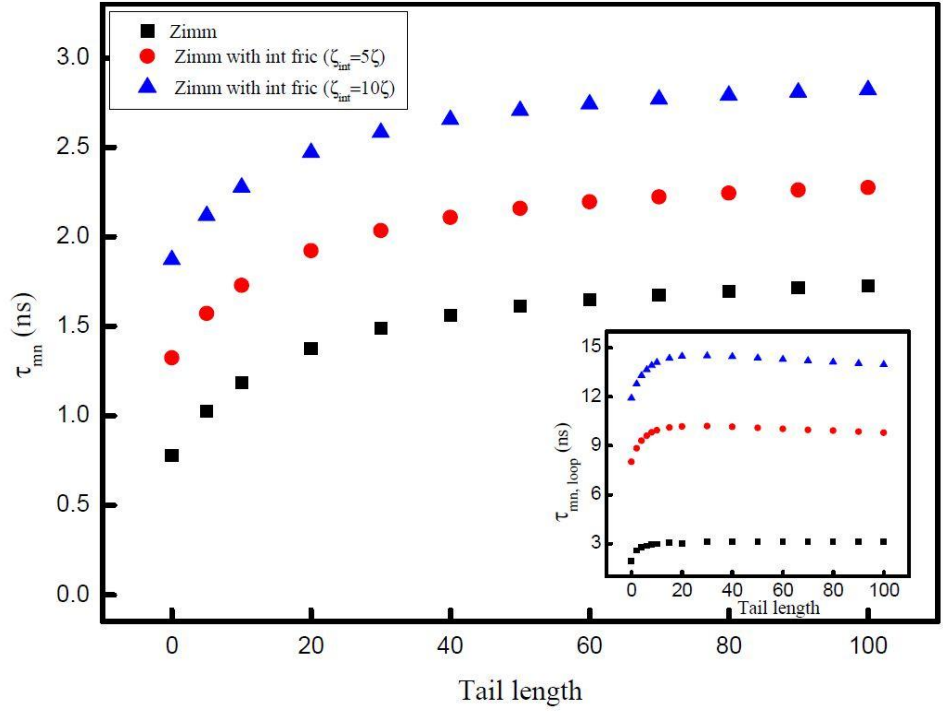

FIG. 13: $\tau_{mn}$ and $\tau_{mn,loop}$ (inset) vs tail length for the Zimm chain and Zimm chain with internal friction, for the topology classes end-to-end and end-to-interior.